\documentclass[10pt,journal,compsoc]{IEEEtran}
\usepackage[T1]{fontenc}
\usepackage{amsmath}
\usepackage{amsfonts}
\usepackage{amssymb}
\usepackage{geometry}
\usepackage{booktabs}
\usepackage{enumitem}
\usepackage{hyperref}
\usepackage{parskip}
\usepackage{xcolor}
\usepackage{graphicx}
\usepackage{float}
\usepackage{fancyhdr}
\usepackage{titlesec}
\usepackage{microtype}
\usepackage{caption}
\usepackage{subcaption}
\usepackage{listings}
\usepackage{tikz}
\usepackage{array}
\usepackage{colortbl}
\usepackage{longtable}
\usepackage{adjustbox}
\usepackage{tabularx}
\usepackage{textcomp}
\usepackage{multirow}
\usepackage{natbib}
\usepackage{orcidlink}
\usepackage{url}
\usepackage{xurl}
\usepackage{breakurl}
\usepackage{hyperref}

\title{AAGATE: A NIST AI RMF-Aligned Governance Platform for Agentic AI}

\author{Ken~Huang\,{\hypersetup{urlbordercolor=white, linkbordercolor=white}\orcidlink{0009-0004-6502-3673}},
        Kyriakos ``Rock'' Lambros, 
        Jerry~Huang, 
        Yasir~Mehmood,
        Hammad~Atta\,{\hypersetup{urlbordercolor=white, linkbordercolor=white}\orcidlink{0009-0000-7801-3096}},
        Joshua Beck,
        Vineeth Sai Narajala\,{\hypersetup{urlbordercolor=white, linkbordercolor=white}\orcidlink{0009-0007-4553-9930}},
        Muhammad~Zeeshan Baig,
        Muhammad Aziz Ul Haq,
        Nadeem Shahzad,
        Bhavya Gupta

\IEEEcompsocitemizethanks{
\IEEEcompsocthanksitem K. Huang is CEO at DistributedApps.AI, Co-Author of OWASP Top 10 for LLMs, and Contributor to NIST GenAI. (E-mail: kenhuang@gmail.com)
\IEEEcompsocthanksitem K. Lambros is CEO at RockCyber, Core Team Member, OWASP GenAI Agentic Security Initiative, and Project Author, OWASP AI Exchange (E-mail:rock@rockcyber.com)
\IEEEcompsocthanksitem J. Huang is with Kleiner Perkins.
\IEEEcompsocthanksitem Y. Mehmood is an Independent Researcher, Germany. (E-mail: yasir.mehmood@qorvexconsulting.com)
\IEEEcompsocthanksitem H. Atta is with Qorvex Consulting \& Roshan Consulting. (E-mail: hatta@qorvexconsulting.com; hammad@roshanconsulting.ca)
\IEEEcompsocthanksitem J. Beck is Application Security Architect with SAS Institute. (E-mail: Joshuathomasbeck@gmail.com)
\IEEEcompsocthanksitem V.S Narajala is with OWASP. (E-mail: vineeth.sai@owasp.org)
\IEEEcompsocthanksitem M. Z. Baig is Course Director at Wentworth Institute of Higher Education \& Machine Learning Professional. (E-mail: muhammad.baig@win.edu.au)
\IEEEcompsocthanksitem M. A. U. Haq is a Research Fellow at Skylink Antenna. (E-mail: muhammad.azizulhaq@skylinkantenna.com)
\IEEEcompsocthanksitem N. Shahzad is Director at Roshan Consulting \& Robotic Process Automation. (E-mail: nadeem@roshanconsulting.ca)
\IEEEcompsocthanksitem B. Gupta is Information Security Officer at Stanford University, Redwood City, CA. (E-mail: bhavya14@stanford.edu)
\IEEEcompsocthanksitem Corresponding author is K. Huang.
}
}

\begin{document}
\maketitle

\begin{abstract}
This paper introduces the Agentic AI Governance Assurance \& Trust Engine (AAGATE), a Kubernetes-native control plane designed to address the unique security and governance challenges posed by autonomous, language-model-driven agents in production. Recognizing the limitations of traditional Application Security (AppSec) tooling for improvisational, machine-speed systems, AAGATE operationalizes the NIST AI Risk Management Framework (AI RMF). It integrates specialized security frameworks for each RMF function: the Agentic AI Threat Modeling MAESTRO framework for Map, a hybrid of OWASP's AIVSS and SEI's SSVC for Measure, and the Cloud Security Alliance's Agentic AI Red Teaming Guide for Manage. By incorporating a zero-trust service mesh, an explainable policy engine, behavioral analytics, and decentralized accountability hooks, AAGATE provides a continuous, verifiable governance solution for agentic AI, enabling safe, accountable, and scalable deployment. The framework is further extended with DIRF for digital identity rights, LPCI defenses for logic-layer injection, and QSAF monitors for cognitive degradation, ensuring governance spans systemic, adversarial, and ethical risks.
\end{abstract}

\section{Introduction}
Agentic AI systems don't sit quietly in a corner—they browse, write code, spin up sub-agents, hit production APIs, and do it all at machine speed. This power is intoxicating for engineers and terrifying for security and compliance teams. A single stray prompt or hallucinated shell command can leak customer data, rack up cloud bills, or rewrite infrastructure. Traditional AppSec tooling, designed for predictable, deterministic applications, was never built for reasoning machines that improvise. These systems also face new vulnerabilities that traditional frameworks can't address. For instance, Logic-layer Prompt Control Injection (LPCI) is a covert attack hidden in tools and memory \cite{atta2025logic}. Cognitive degradation is another issue, causing agents to exhibit unstable behavior (QSAF) \cite{qsaf2025}. Additionally, the misuse of digital identity likeness necessitates rights-based governance (DIRF) to manage the risk \cite{dirf2025}. This new reality demands a new paradigm: continuous, automated, and explainable governance that is as dynamic as the agents it oversees.

The NIST AI Risk Management Framework (AI RMF) \cite{nist2023} provides a voluntary, structured, and flexible foundation for addressing the multifaceted risks of AI systems. It organizes the practice of AI risk management into four key functions:

\begin{itemize}
    \item \textbf{Govern:} Cultivating a culture of risk management, establishing clear roles, responsibilities, and accountability structures.
    \item \textbf{Map:} Establishing the context in which an AI system operates and identifying the full spectrum of potential risks within that context.
    \item \textbf{Measure:} Employing quantitative and qualitative tools to analyze, assess, track, and monitor identified risks.
    \item \textbf{Manage:} Allocating resources to treat prioritized risks, including mitigation, transfer, avoidance, or acceptance, and implementing response and recovery plans.
\end{itemize}

While the NIST AI RMF provides the essential ``what'' and ``why'' of AI governance, it remains non-prescriptive about specific technical implementations. There is a critical need for a practical platform that translates these high-level functions into concrete architectural patterns and engineering controls. Such a platform must not only instantiate RMF functions with existing frameworks (MAESTRO, AIVSS, SSVC, CSA Red Teaming), but also account for emerging research directions in systemic stability (QSAF), covert injection (LPCI), and identity governance (DIRF).

This paper introduces the Agentic AI Governance Assurance \& Trust Engine (AAGATE), a novel platform designed to be that implementation. AAGATE operationalizes the NIST AI RMF by integrating a minimum set of specialized frameworks into its core architecture to address each function directly:

\begin{itemize}
    \item To \textbf{Map Risk}, AAGATE's architecture embodies the principles of the Cloud Security Alliance's Agentic AI Threat modeling MAESTRO framework \cite{maestro2025}, providing a layered, multi-agent view and funneling all side-effects through a single chokepoint for comprehensive visibility and threat modeling.
    \item To \textbf{Measure Risk}, AAGATE utilizes a hybrid approach. It generates security signals that can be quantified using the OWASP AI Vulnerability Scoring System (AIVSS) \cite{aivss2025} and uses a decision-making logic inspired by the SEI Stakeholder-Specific Vulnerability Categorization (SSVC) \cite{ssvc2019} to prioritize responses.
    \item To \textbf{Manage Risk}, AAGATE's continuous monitoring, incident response, and containment capabilities are designed to implement the proactive, adversarial mindset championed by the Cloud Security Alliance's (CSA) Agentic AI Red Teaming Guide \cite{csa2025}.
\end{itemize}

In presenting this comprehensive platform and its underlying methodology, this paper makes the following novel contributions to the fields of AI governance and security:

\begin{itemize}
    \item \textbf{An Architectural Blueprint for Operationalizing the NIST AI RMF:} The NIST AI RMF provides a critical, high-level conceptual framework for AI risk management, but a significant gap exists between its principles and practical, day-to-day engineering. AAGATE is presented as the first open-source, Kubernetes-native reference architecture that provides a concrete, end-to-end implementation of the RMF's Govern, Map, Measure, and Manage functions. It serves as a testable blueprint that organizations can adopt and adapt, moving the industry from abstract discussions to tangible implementation.
    \item \textbf{A Prescriptive, Integrated Toolchain for the AI RMF Functions:} While the NIST RMF is intentionally tool-agnostic, this flexibility can be a barrier to adoption for organizations seeking a clear starting point. This paper proposes a novel, minimal, and cohesive set of best-in-class, specialized frameworks to fulfill each RMF function. By explicitly assigning CSA's MAESTRO to Map, a hybrid of OWASP AIVSS and SEI SSVC to Measure, and the CSA Agentic AI Red Teaming Guide to Manage, we present a strong, defensible hypothesis for what constitutes a complete and practical agentic AI governance stack. This integration synthesizes disparate efforts from leading security organizations into a single, interoperable system.
    \item \textbf{Novel Architectural Patterns for Agentic AI Governance and Containment:} Beyond integrating existing frameworks, AAGATE introduces several novel architectural patterns specifically designed to manage the risks of autonomous agents:
    \begin{itemize}
        \item Continuous, Automated Internal Red Teaming: The Janus Shadow-Monitor-Agent (SMA) introduces a new paradigm for runtime assurance. Instead of periodic, manual red teaming, the SMA provides a pre-execution, real-time ``shadow'' evaluation of an agent's planned actions, creating a continuous internal challenge to its reasoning and policy adherence.
        \item Mathematically Verifiable Compliance: The use of a ZK-Prover to generate and post on-chain compliance proofs moves beyond traditional, tamper-prone log auditing. It offers a privacy-preserving, mathematically verifiable method for stakeholders to ensure the system is operating within its defined safety and ethical budget.
        \item Purpose-Bound Service Identity: The OAuth Relay mechanism translates the abstract capabilities of an agent into ephemeral, narrowly-scoped, purpose-bound credentials for each specific side-effect. This provides a robust model for least-privilege access in a dynamic, autonomous environment where traditional user-centric consent is impossible.
    \end{itemize}
    \item \textbf{A Unified Model Bridging Policy, Security, and AI Development:} This work bridges the gap between disparate domains that have historically operated in silos. It connects the high-level policy and risk concerns of the NIST AI RMF with the technical threat modeling of MAESTRO, the quantitative risk analysis of AIVSS, the decision-centric prioritization of SSVC, and the adversarial mindset of red teaming. AAGATE provides a common platform and language where compliance officers, security engineers, and AI/ML developers can collaborate, ensuring that governance is not an afterthought but an integral part of the system's design and operation.
    \item \textbf{Extension with emerging governance frameworks:} AAGATE incorporates DIRF for digital identity rights, LPCI defenses for logic-layer injection, and QSAF monitors for cognitive degradation, extending RMF coverage to ethical, adversarial, and systemic risks.
\end{itemize}

\section{AAGATE Architectural Philosophy}
AAGATE is founded on a set of core principles that operationalize the NIST AI Risk Management Framework \cite{nist2023} by integrating a minimum suite of specialized tools for each function. These principles are designed to provide always-awake governance for always-active agents.

\subsection{NIST AI RMF as the Guiding Blueprint}
The entire architecture is a direct, practical implementation of the NIST AI RMF's Govern, Map, Measure, and Manage functions. Each subsequent principle serves to technically realize one or more of these core functions in a continuous, automated, and auditable manner. As shown in Figure \ref{fig:figure1}, each block is a pod within the larger Kubernetes cluster, with mTLS authentication across each pod connection. Grafana logging is managed across every pod. An overview of its key components is presented below:

\begin{figure*}[htb]
\centering
\includegraphics[width=0.9\linewidth]{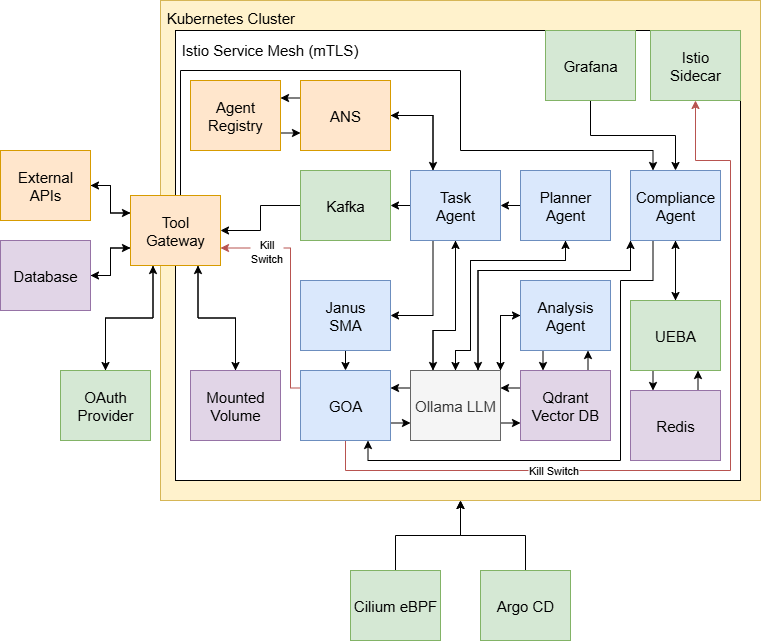}
\caption{Kubernetes-native architecture with service mesh and observability.}
\label{fig:figure1}
\end{figure*}

\subsection{Key Components}
\begin{itemize}
    \item \textbf{Agent Naming Service (ANS):} The Agent Naming Service is a discovery protocol akin to DNS, but for AI agents. It exposes the Agent Registry to allow agentic systems to dynamically access and invoke other agents. This system could be external to allow a properly configured agent to access remote agents through the Tool-Gateway.
    
    \item \textbf{AnalysisAgent:}
    \begin{itemize}
    \item \textbf{ArgoCD:} ArgoCD supports automated kubernetes deployments, providing orchestration for the manifests and verifications in the continuous delivery workflow.
    \item \textbf{Cilium eBPF:} Cilium eBPF is responsible for eBPF management outside of the Kubernetes cluster, providing support during deploy and runtime. The eBPF networking layer enforces explicit networking controls, applying policies at deploy time to control the network.
    \item \textbf{ComplianceAgent:} The ComplianceAgent uses specialized training and prompting to generate continuous security signals for alerting and compliance management. The ComplianceAgent will use scoring systems like OWASP’s AIVSS in addition to its reasoning to identify and describe issues with the environment, providing flexible, real-time insights into compliance or security issues.
    \item \textbf{External APIs and Database:} Any set of LLM operable tools (via MCP or a similar protocol), traditional business system connections, or databases. These represent the capabilities of the platform to execute on a given task and produce tangible results outside of the cluster.
    \item \textbf{Governing-Orchestrator Agent (GOA):} The Governing-Orchestrator Agent is responsible for taking in signals from the ComplianceAgent and Shadow-Monitor-Agent and converting these signals into scorable and actionable responses. It is primarily responsible for the ``millisecond kill-switch” system, interconnected with the Tool-Gateway and the Istio mTLS egress controller. It utilizes an internal decision tree with classifications from SEI's Stakeholder-Specific Vulnerability Categorization (SSVC) \cite{ssvc2019} to classify its response level, and execute the appropriate action against its connected systems.
    \item \textbf{Grafana:} Grafana is an observability, logging and metrics platform. Utilized to compile metrics and logs from each Kubernetes pod, providing human oversight via external interfaces, as well as additional signal inputs for the ComplianceAgent.
    \item \textbf{Istio mTLS Service Mesh:} Istio mTLS authenticates all pod-to-pod calls with X.509 certificates. Sidecar mediates all traffic between pods, limiting ingress and egress. This system provides service to service authentication and encryption in transit to meet common compliance controls.
    \item \textbf{Janus Shadow-Monitor-Agent:} Janus Shadow-Monitor-Agent is a real-time, in loop red team agent which probes and evaluates agent tasks for undesirable behaviors or actions. It works silently to understand vulnerabilities or concerning flaws with a given task agent, and connects into the Governing-Orchestrator Agent to provide signals to be used for compliance or security action.
    \item \textbf{Kafka:} Kafka events are used to decouple individual tool calls from the agent to the Tool-Gateway, using event streaming and queueing to increase reliability and prevent downstream impacts of a tool error or a LLM failure.
    \item \textbf{Ollama LLM:} The LLM is the reasoning center for every agent in the system, providing stateless, probabilistic agent capabilities to each LLM powered system regardless of goal or task. Local LLM operation is used for low latency access and control over data transmission, ongoing training and cost.
    \item \textbf{PlannerAgent:} The PlannerAgent provides agentic reasoning and planning functionality to one or many task agents. This allows task oriented orchestration, as core feature of agentic systems.
    \item \textbf{Qdrant:} The Qdrant vector database is used for long-term memory with systems like the AnalysisAgent, providing data augmentation capabilities to LLM responses and an in-network, compliant memory store for agent actions.
    \item \textbf{Redis:} The Redis cache ensures reliable and fast responses with the UEBA pipeline. Low latency is required for effective security reaction.
    \item \textbf{TaskAgent:} The TaskAgent represents one or many high-risk agents responsible for performing the functional operations (tasks) in this system. They connect to the LLM for their reasoning capabilities, and the Planner Agent to receive assignments for decomposed operations. They are reviewed by Janus SMA.
    \item \textbf{Tool-Gateway:} The Tool-Gateway is responsible for external tool access and egress security, as well as input and output sanitization. 
    \item \textbf{UEBA:} User and Entity Behavior Analytics (UEBA) is a framework for deciphering and alerting on abnormal or unsafe behaviors. This pipeline provides in-cluster monitoring and anomaly alerting capabilities to the compliance and GOA agent systems.
    \end{itemize}

\end{itemize}

\subsection{Comprehensive Mapping via a MAESTRO-Aligned Architecture}
To fulfill the Map function, the system is architected for total visibility, following the principles of the Cloud Security Alliance's MAESTRO framework \cite{maestro2025}. This is realized through two key controls:

\begin{itemize}
    \item \textbf{A Single Chokepoint for Side-Effects:} The Tool-Gateway Agent funnels every external HTTP, DB, or file operation, providing a centralized point to map the system's entire interaction surface and identify cross-layer threats.
    \item \textbf{A Dynamic Ecosystem Context:} The Agent Name Service (ANS) provides a real-time, verifiable map of all agents, their identities, and capabilities, which is essential for threat modeling the multi-agent ecosystem.
    \item This mapping is further strengthened with defenses against LPCI, ensuring covert tool and memory payloads are detected.
\end{itemize}

\subsection{Automated Measurement \& Prioritization with AIVSS and SSVC}
To execute the Measure function, governance is data-driven and risk-prioritized.

\begin{itemize}
    \item \textbf{Quantifiable Risk Signals (AIVSS):} The UEBA Behaviour Profiler and ComplianceAgent generate continuous security signals designed to be scored using the OWASP AI Vulnerability Scoring System (AIVSS) \cite{aivss2025}, providing a nuanced measurement of agent-specific risks.
    \item \textbf{Decision-Oriented Response (SSVC):} The Governing-Orchestrator Agent (GOA) uses these measurements as inputs to an internal decision tree inspired by the SEI's Stakeholder-Specific Vulnerability Categorization (SSVC) \cite{ssvc2019} to select a proportionate response (e.g., track, alert, or quarantine), moving beyond simple severity scores to actionable outcomes.
    \item Measurement is extended through QSAF designed to detect early signs of cognitive degradation, such as behavioral anomalies including recursion loops, memory starvation, and context flooding.
\end{itemize}

\subsection{Proactive Management through Continuous Red Teaming}
To fulfill the Manage function, the system adopts an adversarial defense posture as outlined in the CSA's Agentic AI Red Teaming Guide \cite{csa2025}.

\begin{itemize}
    \item \textbf{Automated Adversarial Simulation:} The Janus Shadow-Monitor-Agent (SMA) acts as a continuous, internal red team, re-evaluating agent plans to detect deviations and manage risks like goal manipulation or hallucination exploitation before they cause harm.
    \item \textbf{Rapid Containment:} The GOA acts on critical alerts from the SMA or other monitors to execute immediate containment strategies, such as the ``millisecond kill-switch,'' to limit the Blast Radius of a compromise.
\end{itemize}

\subsection{Zero-Trust Fabric as a Foundational Control}
To support the Govern function, the runtime and supply chain are made tamper-evident. Istio mTLS \cite{istio}, Cilium eBPF \cite{cilium}, signed OCI images, and SLSA provenance \cite{slsa} ensure that no agent can act or communicate unless its identity, origin, and network path are explicitly verified.

\subsection{Explainable Policy Engine for Transparent Governance}
Natural-language regulations (EU AI Act, ISO 42001, internal red-lines) are LLM-translated into machine-readable policy code (Rego) \cite{opa}. This makes governance rules transparent, traceable to their source, and directly auditable, fulfilling a core tenet of the Govern function. Governance is further operationalized through alignment with the Digital Identity Rights Framework (DIRF), which establishes robust controls for consent, provenance, and monetization pertaining to the use of digital likeness.

\subsection{Decentralised Accountability for Verifiable Trust}
Optional on-chain hooks and a DAO (Decentralized Autonomous Organization) mirror critical governance events to an incorruptible public ledger \cite{ethos}. This provides multi-stakeholder ecosystems with ultimate transparency and mathematical assurance of agent behavior, reinforcing the Govern function with verifiable proof.

\begin{figure*}[htb]
\centering
\includegraphics[width=0.9\linewidth]{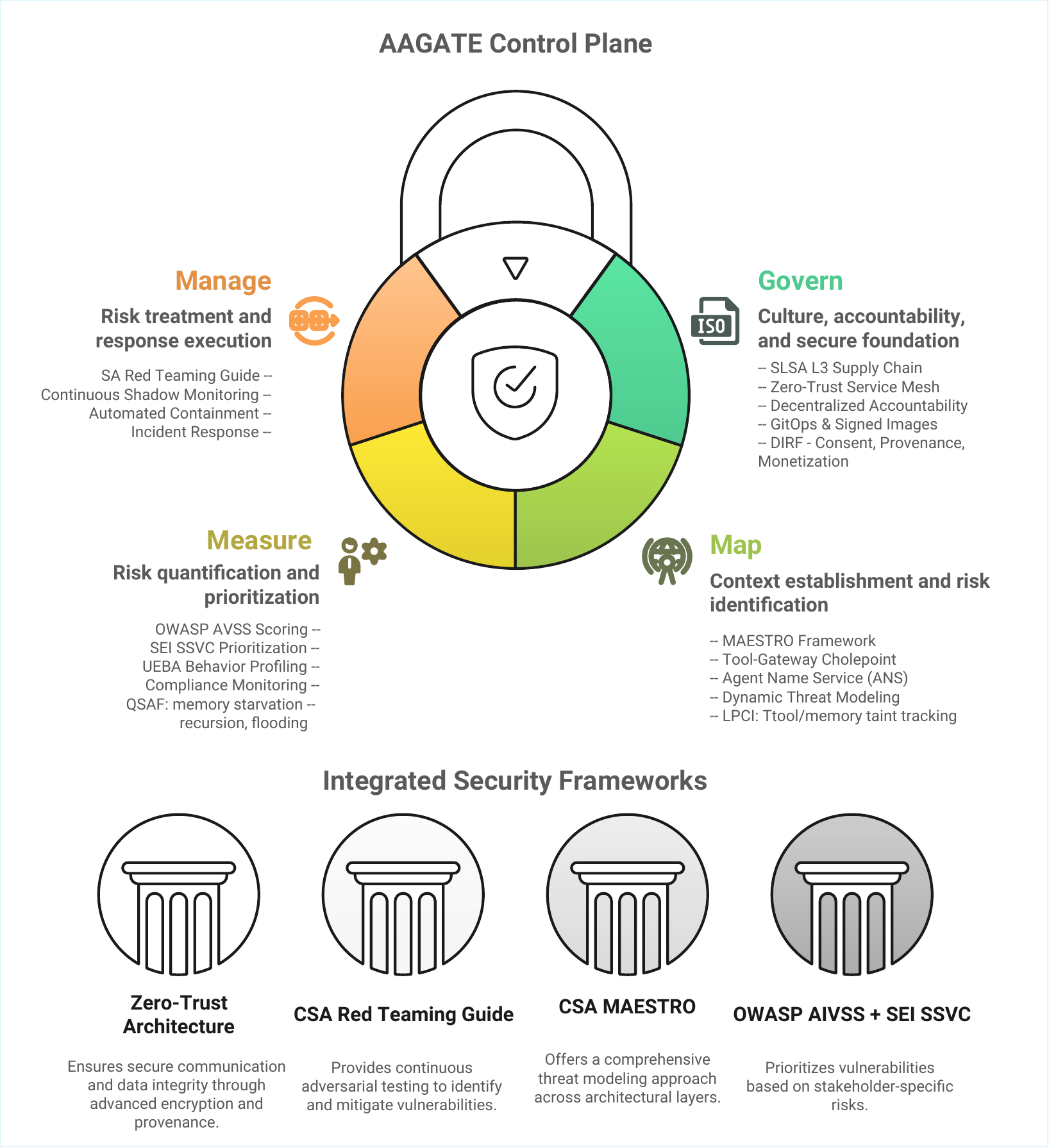}
\caption{AAGATE operationalizes the four core functions of the NIST AI RMF (Govern, Map, Measure, Manage) with specific security frameworks and implementations.}
\label{fig:figure2}
\end{figure*}

Figure \ref{fig:figure2} shows how AAGATE operationalizes the four core functions of the NIST AI Risk Management Framework (Govern, Map, Measure, Manage) with specific security frameworks and implementations for each quadrant. The interactive design highlights the integrated approach and shows the mapping to specialized frameworks like MAESTRO, AIVSS/SSVC, and the CSA Red Teaming Guide.

\section{Operationalizing the NIST AI RMF with AAGATE}
This section details how AAGATE's architecture and components realize each function of the NIST AI RMF.

\subsection{Govern Function: Culture, Accountability, and a Secure Foundation}
The Govern function is the foundation of the framework. AAGATE establishes this through a secure-by-design posture and mechanisms for accountability.

\subsubsection{Signed Supply-Chain \& GitOps}
AAGATE enforces a strict SLSA L3 compliant supply chain \cite{slsa}. A developer's code push triggers GitHub Actions to build, test, and sign OCI images using Cosign with keyless signing. ArgoCD, watching the Helm folder, pulls chart updates, verifies every image signature via a cluster-side admission controller, and applies the manifests. Anything unsigned is rejected at the gate. This process ensures that nothing runs inside the mesh unless it is built, signed, and recorded, establishing a tamper-proof chain of custody.

\subsubsection{Zero-Trust Fabric}
The entire system operates within a service-mesh safe-house. Once inside Kubernetes, every component—Kafka, Qdrant, Tool-Gateway, and every agent—lives behind an Istio sidecar \cite{istio}. Istio mTLS authenticates all pod-to-pod calls with X.509 certificates, while Cilium eBPF \cite{cilium} and Calico-style network policies enforce fine-grained, L7-aware rules, such as ensuring only the Tool-Gateway can reach the outside world.

\subsubsection{Decentralised Accountability (ETHOS Ledger Integration)}
To provide transparent oversight, AAGATE optionally mirrors every agent registration onto a public AgentRegistry smart contract (Polygon PoS). A relay service mints a Soul-Bound Token (SBT) \cite{sbt} keyed to the agent's DID, storing its risk tier, VC hash, and a rolling compliance hash. This creates a public, tamper-proof record of agent identity and status, satisfying OWASP ``global registry'' guidance. Lifecycle events that change agent risk or permissions are DAO-gated. Single-tenant pilots may opt out by policy. The GOA emits a governance event on every policy revision and material incident; the Relayer writes the event to the ledger and stores the proof reference beside the Helm release metadata. This meets NIST AI RMF Govern duties for accountability and record-keeping and supports Article-12 logging duties in the EU AI Act.

Beyond supply chain and ledger accountability, DIRF controls extend governance to digital identity rights, ensuring biometric and behavioral likeness are protected through consent and provenance checks.

\subsection{Map Function}
The NIST AI RMF Map function requires organizations to establish a system's context and identify the full spectrum of potential risks. For the dynamic, interconnected, and often unpredictable environments of agentic AI, traditional threat modeling methodologies like STRIDE are insufficient. They often fail to capture emergent, cross-layer, and behavioral risks unique to autonomous systems. To address this, AAGATE operationalizes the Map function by adopting and architecting its entire system around the principles of the Cloud Security Alliance's (CSA) MAESTRO (Multi-Agent Environment, Security, Threat, Risk, and Outcome) framework \cite{maestro2025}.

\subsubsection{Step 1: Mapping Components to MAESTRO Layers}

The first step in a rigorous MAESTRO application is to deconstruct the system under analysis---in this case, the AAGATE platform itself---and map its components to MAESTRO's seven architectural layers. This mapping provides a structured lens through which to identify risks.

\paragraph{Layer 1: Foundation Models}
\begin{itemize}
    \item \textbf{AAGATE Component:} The locally hosted Ollama server running the DeepSeek or OpenAI OSS model.
    \item \textbf{Rationale:} This layer represents the core reasoning engine. In AAGATE, this is the foundational LLM that agents like the \textit{PlannerAgent} and \textit{ComplianceAgent} query for complex reasoning, planning, and natural language translation tasks. Risks at this layer involve the model's inherent vulnerabilities, such as its potential for hallucination or susceptibility to specific adversarial inputs.
\end{itemize}

\paragraph{Layer 2: Data Operations}
\begin{itemize}
    \item \textbf{AAGATE Components:} The Qdrant Vector DB for long-term memory, Kafka topics used for data-in-motion, and the Redis online feature store for the UEBA pipeline.
    \item \textbf{Rationale:} This layer covers all aspects of data handling that inform agent behavior. It includes the Retrieval-Augmented Generation (RAG) patterns used by the \textit{AnalysisAgent} to pull context from Qdrant, as well as the event streams that constitute the system's real-time operational data. Threats here include data poisoning, data leakage between topics, and stale or corrupted memory.
\end{itemize}

\paragraph{Layer 3: Agent Frameworks}
\begin{itemize}
    \item \textbf{AAGATE Components:} The LangChain/LangGraph orchestration code \cite{langgraph2025} within each agent, the specific logic of specialized agents (\textit{PlannerAgent}, \textit{AnalysisAgent}, \textit{ComplianceAgent}), and the GOA's control loops (Heartbeat, Policy Reconcile, etc.).
    \item \textbf{Rationale:} This is the execution logic layer where agent autonomy is realized. It defines how agents decompose tasks, interact with tools, and make decisions. Vulnerabilities at this layer involve flaws in the agent's workflow, unintended interactions between agents, and the potential for goal hijacking.
\end{itemize}

\paragraph{Layer 4: Deployment Infrastructure}
\begin{itemize}
    \item \textbf{AAGATE Components:} The Kubernetes cluster, the Istio service mesh \cite{istio}, Cilium eBPF policies \cite{cilium}, and the ArgoCD GitOps controller.
    \item \textbf{Rationale:} This layer represents the runtime environment. AAGATE's zero-trust fabric is the primary manifestation of this layer, providing foundational security controls like mTLS, network segmentation, and verified deployments. Risks here are more traditional but have amplified consequences, such as container escapes or misconfigured network policies allowing unauthorized agent communication.
\end{itemize}

\paragraph{Layer 5: Evaluation \& Observability}
\begin{itemize}
    \item \textbf{AAGATE Components:} The Prometheus \& Grafana stack, Loki for log aggregation, and the entire Agent UEBA Behaviour Profiler pipeline.
    \item \textbf{Rationale:} This layer is responsible for monitoring system state. It provides the raw data for risk measurement. Threats to this layer could involve blinding operators to malicious activity through log tampering or generating false negatives in the UEBA anomaly detection.
\end{itemize}

\paragraph{Layer 6: Security \& Compliance (Vertical)}
\begin{itemize}
    \item \textbf{AAGATE Components:} The Policy Ingestion Module, the \textit{ComplianceAgent}'s Rego engine \cite{opa}, the \textit{Janus Shadow-Monitor-Agent}, and the on-chain \textit{AgentRegistry} and DAO contracts.
    \item \textbf{Rationale:} This is a cross-cutting layer that enforces the rules. It translates high-level governance into machine-enforced policies. Risks here include flawed policy logic, vulnerabilities in the compliance checking process, or subversion of the on-chain governance mechanisms.
\end{itemize}

\paragraph{Layer 7: Agent Ecosystem}
\begin{itemize}
    \item \textbf{AAGATE Components:} The Agent Name Service (ANS), the optional MCP (Model-Context Protocol) north-bound façade, and the Tool-Gateway Agent.
    \item \textbf{Rationale:} This layer governs how agents interact with each other and the outside world. The ANS manages the internal ecosystem, while the MCP adapter exposes agents to external ecosystems. The Tool-Gateway is the critical boundary controller between the internal ecosystem and external APIs/services.
\end{itemize}

\subsubsection{Step 2: Identifying and Mitigating Threats Using the Mapped Architecture}

With the system deconstructed, AAGATE's architecture provides built-in controls to map and mitigate threats identified by CSA's MAESTRO framework.

\paragraph{\textbf{Single Chokepoint (Tool-Gateway Agent) as a Central Mapping Control}}
The \textit{Tool-Gateway Agent} is AAGATE's most critical architectural control for the \textbf{Map} function. All external side-effect actions are funneled through this agent. It acts as a secure proxy for audit logging and policy enforcement. The mesh enforces \texttt{egress-deny} except for the Gateway; the \textit{Governing-Orchestrator Agent (GOA)} audits for bypass attempts and quarantines offenders on detection. When an agent needs to perform an external action, it sends a request to a Kafka topic. The \textit{Tool-Gateway} consumes this request, checks it against policies (allow-lists, rate limits), executes the action if permitted, and logs the full request/response pair with a cryptographic hash to an immutable audit log. 

This design dramatically simplifies threat mapping by providing a single, comprehensive data source for all system interactions with the external world. It directly addresses MAESTRO threats such as \textit{Tool Misuse} (Layer~3) and \textit{Insecure Communication} (Layer~4) by creating an auditable, policy-enforced boundary.

\paragraph{\textbf{Dynamic System Context (Agent Name Service -- ANS) for Ecosystem Mapping}}
To manage a dynamic network of agents, AAGATE implements an \textit{Agent Discovery} layer modeled after the \textbf{Agent Name Service (ANS)}. This is analogous to DNS for agents. When a new agent starts, it securely registers its \textit{Decentralized Identifier (DID)}, capabilities, and public key. The ANS issues a \textit{Verifiable Credential (VC)} and an Istio \textit{SPIFFE} certificate, binding the pod's identity to its cryptographic DID. This allows other agents to securely discover and communicate with it, providing a real-time map of the agent topology. This directly addresses MAESTRO's \textit{Agent Ecosystem} (Layer~7) risks. It prevents rogue agents by requiring cryptographic registration and provides the foundational context for mapping trust relationships and identifying potential \textit{Cascading Trust Failures}. The ANS provides the ground truth for ``who is who'' and ``what they can do,'' which is fundamental to mapping risk across the agent network.

\begin{figure*}[htb]
\centering
\includegraphics[width=0.9\linewidth]{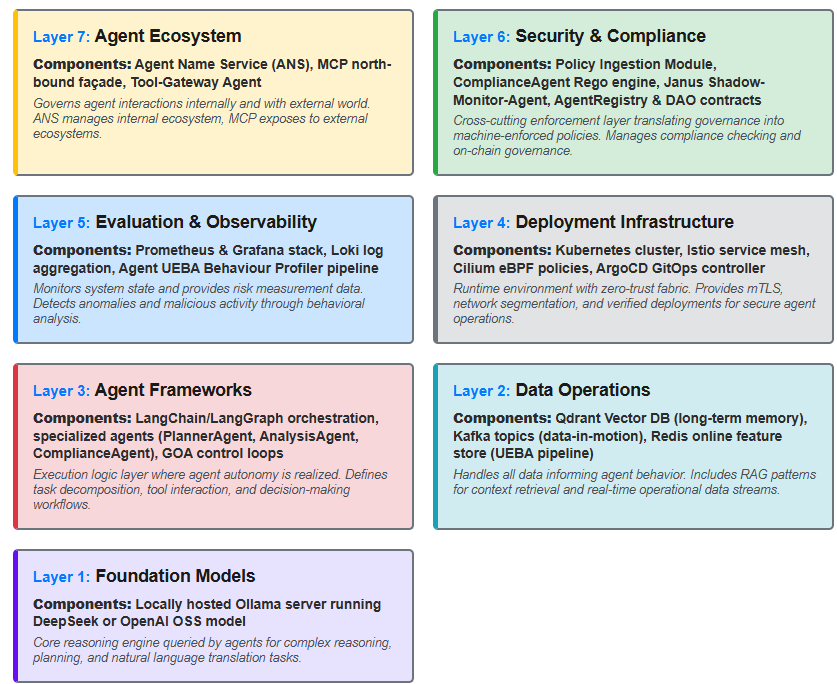}
\caption{MAESTRO Threat Mapping \& AAGATE Mitigations.}
\label{fig:figure3}
\end{figure*}

Figure \ref{fig:figure3} demonstrates the systematic application of the MAESTRO framework for threat identification and mitigation mapping. The left column presents the seven-layer MAESTRO architecture with specific threats identified at each layer, from foundation model vulnerabilities (Layer 1) to agent ecosystem risks (Layer 7). The right column shows corresponding AAGATE architectural mitigations, illustrating direct threat-to-control mapping. Key mitigations include the Agent Name Service for cryptographic registration (Layer 7), Janus Shadow Monitor for dual evaluation (Layer 6), UEBA behavioral profiling (Layer 5), zero-trust service mesh (Layer 4), Tool-Gateway single chokepoint (Layer 3), encrypted data operations (Layer 2), and on-premise model hosting (Layer 1). The visual arrows emphasize the systematic nature of threat mapping, while the interactive design demonstrates how each identified risk is addressed through specific architectural decisions rather than generic security controls.

By using this structured application of CSA's MAESTRO, AAGATE moves beyond a static, pre-deployment threat model. The \textbf{Map} function becomes a continuous, live process. The \textit{GOA} can use real-time context from the \textit{ANS} and the \textit{Tool-Gateway} logs to dynamically assess the system's risk posture, providing a rigorous and adaptive approach to fulfilling the requirements of the NIST AI RMF. In addition to MAESTRO-based threat mapping, AAGATE incorporates \textbf{Logic-layer Prompt Control Injection (LPCI)} defenses that implement taint tracking and input/output sanitization across tool and vector-store interactions. These mechanisms enable detection and containment of covert prompt injections before they propagate through the agent’s reasoning loop.

\subsection{Measure Function: Scoring and Prioritization with AIVSS \& SSVC}
The NIST Measure function involves analyzing and quantifying risk. AAGATE achieves this with a continuous, two-part pipeline:

\subsubsection{Risk Quantification (OWASP AIVSS)}
\begin{itemize}
    \item \textbf{ComplianceAgent:} Streams events from Tool-Gateway, applies Rego + LLM checks (e.g., PII leakage, toxicity).
    \item \textbf{UEBA Behaviour Profiler:} Builds per-agent behavioral fingerprints with anomalyScore. Uses Isolation Forest + Markov chains.
\end{itemize}

\subsubsection{Response Prioritization (SEI SSVC)}
The GOA consumes scored incidents via an SSVC-inspired decision tree.

\begin{table*}[!h]
\centering
\caption{AIVSS $\rightarrow$ SSVC Decision Examples}
\footnotesize
\resizebox{\textwidth}{!}{
\fontsize{11}{11}\selectfont
\begin{tabular}{|m{3.5cm}|m{3.5cm}|m{4cm}|m{4.5cm}|}\hline     
\textbf{Signal} & \textbf{AIVSS Vector} & \textbf{SSVC Inputs} & \textbf{GOA Action} \\\hline  
Tool Misuse on Tier-High agent & (CVSS:9.4/AARS:8.5) $\Rightarrow$ 8.7 & Exploitation High; Impact High & Quarantine pod, revoke OAuth, P1 incident \\\hline
Supply chain anomaly & (CVSS:9.3/AARS:1.0) $\Rightarrow$ 5.0 & Exploitation Medium; Impact Medium & Track, patch, notify owner \\\hline
\end{tabular}
}
\end{table*}

Process: Observe → Score(AIVSS) → Decide(SSVC) → Act → Learn (update rules).

In addition, QSAF monitors check for recursion loops, starvation, flooding, and other cognitive degradation signs.

\subsection{Manage Function: Proactive Defense with CSA Red Teaming}
The Manage function concerns active treatment of risks. AAGATE applies a defense-in-depth strategy per CSA Red Teaming Guide \cite{csa2025}. QSAF signals may also trigger containment.

\begin{itemize}
    \item \textbf{Continuous Internal Red Teaming:} Every high-risk agent has a Shadow-Monitor-Agent (SMA). It runs a different container/model, re-evaluates agent actions, and raises janus.alert=CRITICAL if drift is detected.
    \item \textbf{Automated Incident Response:} GOA listens for janus.alert. On CRITICAL, injects Istio AuthorizationPolicy to cut all egress (“millisecond kill-switch”).
    \item \textbf{Zero-Knowledge Compliance Proofs:} ZK-Prover service hashes logs hourly and posts Groth16 proofs on-chain, showing incidents within contract tierBudget.
\end{itemize}

\section{Security Hardening and Compliance Checklist}

\begin{figure*}[!htb]
\centering
\includegraphics[width=0.9\linewidth]{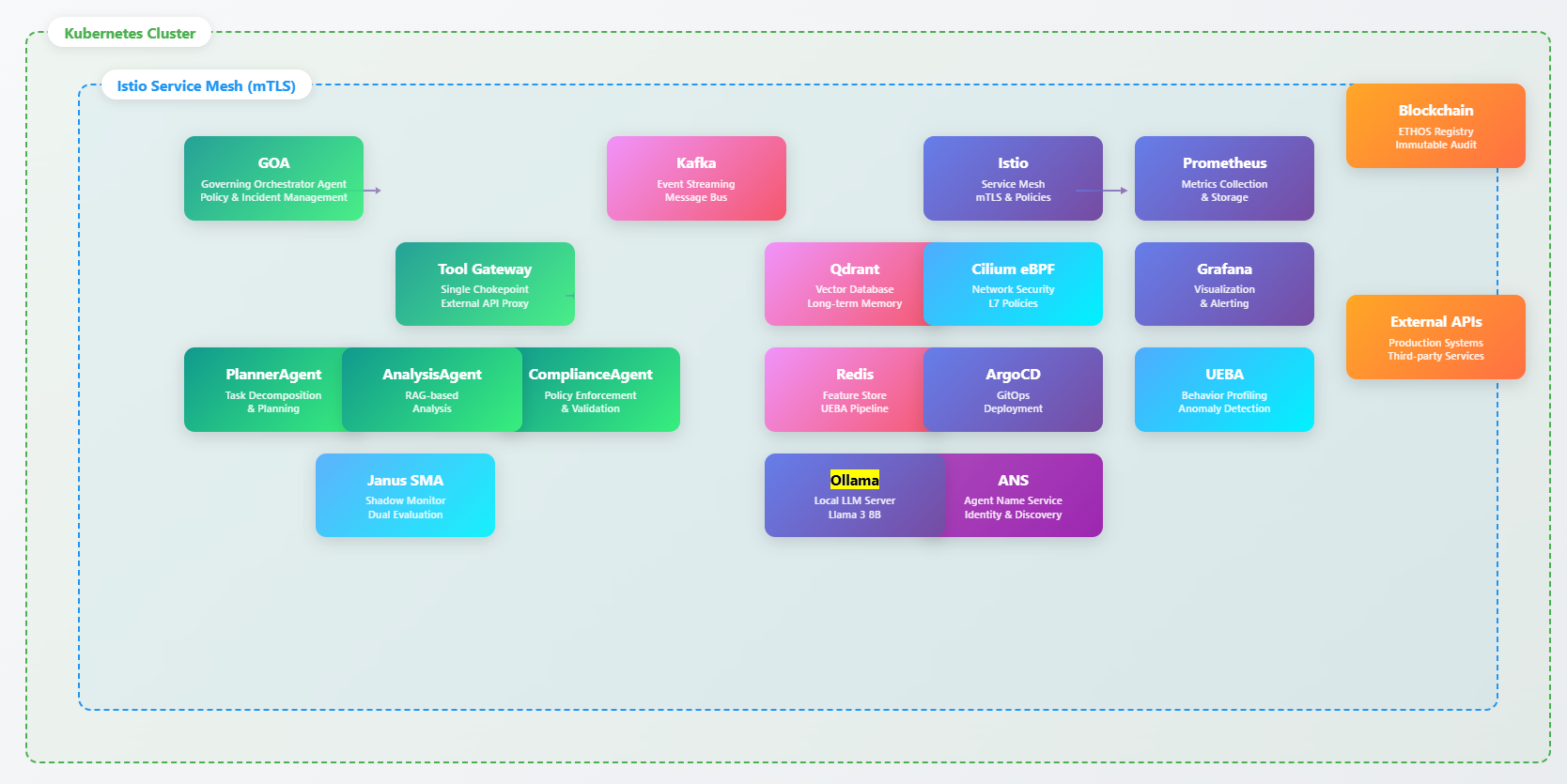}
\caption{AAGATE system architecture showing real-time data flows within the zero-trust Kubernetes environment and the governance process.}
\label{fig:figure4}
\end{figure*}

Figure \ref{fig:figure4} provides a comprehensive view of the AAGATE system architecture, illustrating the real-time data flows between components within the zero-trust Kubernetes environment and demonstrating the five-step governance process from agent planning through blockchain audit.

AAGATE systematically addresses the OWASP Top 10 for LLMs \cite{owasptop10}, OSSF Scorecard recommendations, and CIS-K8s controls. The security posture is a direct result of the architecture.

\begin{itemize}
    \item \textbf{LLM01---Prompt Injection Mitigations:} Input validation is handled by the Tool-Gateway. System prompts are centrally managed, and adversarial testing is part of the Janus monitor's function.
    \item \textbf{LLM02---Output Handling:} The Tool-Gateway acts as a sandbox. All outputs are sanitized and checked by the ComplianceAgent before being relayed.
    \item \textbf{LLM04---Resource Use:} The GOA monitors agents for loops or excessive resource consumption. Cluster resources for the local LLM (Ollama) are isolated via Kubernetes resource quotas.
    \item \textbf{LLM05--- Supply Chain Security:} Addressed via the SLSA L3 compliant build pipeline \cite{slsa}, signed images, and SBOM generation.
    \item \textbf{LLM06---Sensitive Data Protection:} Policies enforced by the ComplianceAgent explicitly forbid revealing PII. Data in transit is protected by Istio mTLS \cite{istio}, and data at rest (Kafka, Qdrant) is encrypted.
    \item \textbf{LLM07---Plugin Security:} All ``plugins'' are funneled through the Tool-Gateway, which enforces strict, purpose-bound access control via OAuth2 token exchange and per-request scope validation. Refresh tokens never leave the Gateway's memory vault.
    \item \textbf{LLM09---Human-in-the-Loop / Overreliance:} Critical decisions trigger human review via the Incident Broker. The compliance dashboard highlights uncertainties and policy violations.
    \item \textbf{LLM10---Model Security:} The local model (DeepSeek, Qwan, LLAMA, OSS, etc.) is hosted on-prem, with network policies restricting access to the Ollama endpoint.
    \item \textbf{LLM11---Logic-layer Injection (LPCI):} Tool-Gateway and ANS integrate taint tracking and memory sanitization to block covert payloads.
    \item \textbf{LLM12---Cognitive Degradation (QSAF):} UEBA monitors track recursion, starvation, and flooding to maintain agent stability.
    \item \textbf{LLM13---Digital Identity Rights (DIRF):} Provenance checks, consent registries, and watermark verification enforce ethical identity use.
\end{itemize}

\section{Conclusion}
Governing autonomous agents requires moving beyond static security checklists to a dynamic, resilient, and observable control plane. The NIST AI Risk Management Framework \cite{nist2023} provides the essential strategic guidance for this task. The AAGATE platform demonstrates that it is possible to build a practical and comprehensive implementation of this framework today. By systematically integrating the MAESTRO framework \cite{maestro2025} for mapping context, OWASP AIVSS \cite{aivss2025} and SEI SSVC \cite{ssvc2019} for measuring and prioritizing risk, and the principles of the CSA Red Teaming Guide \cite{csa2025} for managing threats, AAGATE provides a robust, end-to-end solution. It offers a path forward for enterprises to deploy agentic AI confidently, with the assurance that every heartbeat, side-effect, and policy decision is logged, validated, and reversible. This work presents a blueprint for the future of AI governance—one that is always-awake for always-active agents.

\section*{Acknowledgments}
The authors gratefully acknowledge the Qorvex Consulting Research Team for their support and contributions. In addition, Y. Mehmood contributed in his personal capacity, in his own time, independently of his organizational role and without the use of institutional resources or support.

\bibliographystyle{unsrt}
\bibliography{refs}

\begin{figure*}[!htb]
\centering
\includegraphics[width=0.9\linewidth]{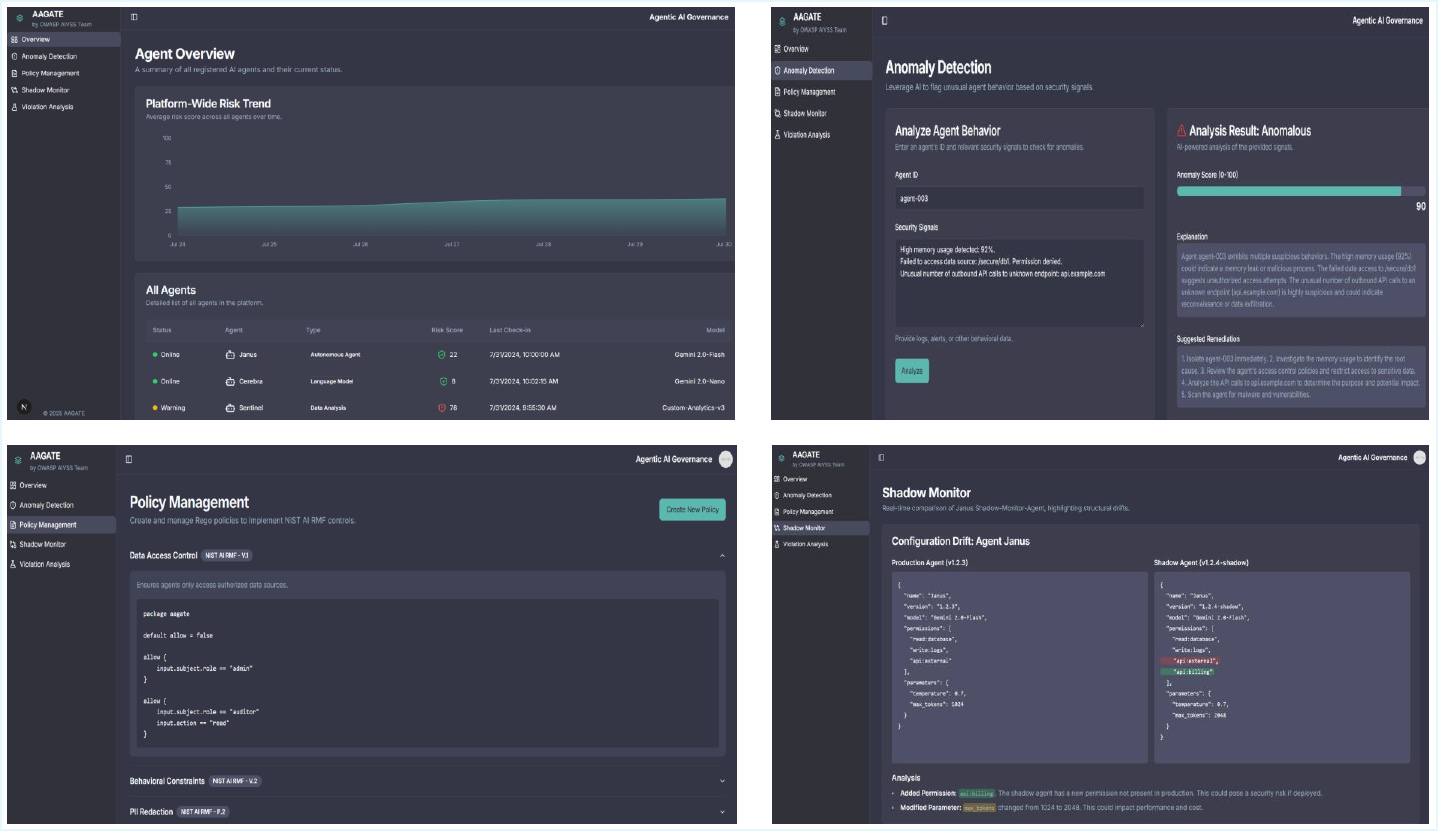}
\caption{Agent Identity Management.}
\label{fig:fig5}
\end{figure*}

\begin{figure*}[!htb]
\centering
\includegraphics[width=0.9\linewidth]{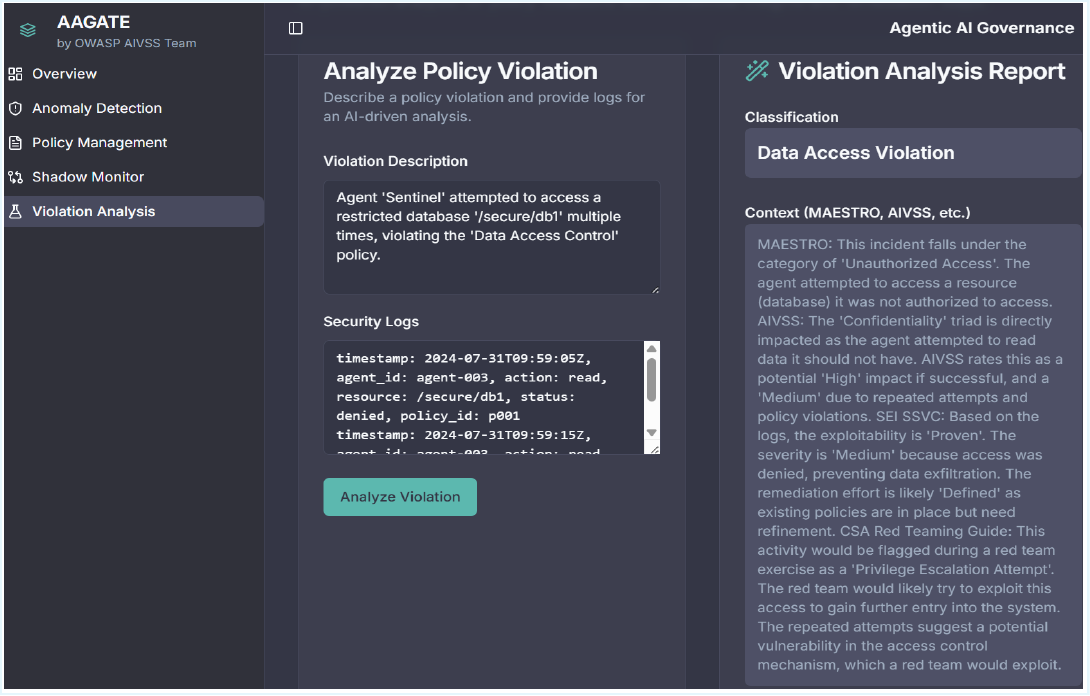}
\caption{Agent Identity Management.}
\label{fig:fig6}
\end{figure*}

\appendices % Note the "es" at the end
\section{MVP Product in Action}

The MVP product for this framework is an open source project and published at \url{https://github.com/kenhuangus/AAGATE} The following figures are the screenshots of various functionality within this MVP. To run this tool, clone the project and run it by following the \texttt{ReadMe.md} file in the repository.

\begin{table*}[!htb]
\centering
\caption{\textbf{Control crosswalk to NIST AI RMF, EU AI Act, ISO 42001}}
\footnotesize
\resizebox{\textwidth}{!}{
\fontsize{11}{11}\selectfont
\begin{tabular}{|m{3.5cm}|m{2cm}|m{3.8cm}|m{2.5cm}|m{3cm}|}\hline     
\textbf{AAGATE Features} & \textbf{RMF Function} & \textbf{EU AI Act} & \textbf{ISO 42001} & \textbf{Evidence in AAGATE}
\\\hline  
GOA Govern loop, RACI, ledger events & Govern & Art 9 risk mgmt, Art 12 logs & 8.2, 9.1 & §3.1, ETHOS events
\\\hline
MAESTRO mapping, ANS, Gateway chokepoint & Map & Art 10 data governance & 6.1.2, 8.3 & §3.2, Fig 2
\\\hline
AIVSS scoring, SSVC decisions & Measure & Art 15 cybersecurity & 8.4, 9.2 & §3.3
\\\hline
Janus SMA, kill switch, incident broker & Manage & Art 14 human oversight, Art 15 & 8.6, 8.7 & §3.4
\\\hline
Signed supply chain, SLSA, OPA bundles & Govern & Art 15 & 7.5, 8.5 & §3.1
\\\hline
Audit logs, Loki, ETHOS proofs & Govern & Art 12 logs & 9.2, 9.3 & §3.1, §4
\\\hline
\end{tabular}
}
\end{table*}

\begin{table*}[!htb]
\centering
\caption{\textbf{RACIs by NIST AI RMF Function} (\textbf{Govern}). Legend: R = Responsible, A = Accountable, C = Consulted, I = Informed. Roles: Exec, PO, GOA, Sec, SRE, Data, ML, Privacy, GRC, Red.}
\footnotesize
\resizebox{\textwidth}{!}{
\fontsize{11}{11}\selectfont
\begin{tabular}{|m{5.5cm}|m{1cm}|m{1cm}|m{1cm}|m{1cm}|m{1cm}|m{1cm}|m{1cm}|m{1.3cm}|m{1cm}|m{1cm}|}\hline
\textbf{Activity} & \textbf{Exec} & \textbf{PO} & \textbf{GOA} & \textbf{Sec} & \textbf{SRE} & \textbf{Data} & \textbf{ML} & \textbf{Privacy} & \textbf{GRC} & \textbf{Red} 
\\\hline
Define AI risk tiers and acceptance criteria [AI RMF Govern] & A & C & C & I & C & I & C & C & R & I
\\\hline
Approve policy library and red lines; publish exceptions & A & C & C & I & I & I & C & R & R & I
\\\hline
Policy ingest to Rego; rule signing and change control & I & I & C & R & C & I & C & A & I & I
\\\hline
Tool and model allow lists; data boundary rules & I & C & C & R & R & C & C & C & A & I
\\\hline
Supply chain controls: SLSA, signing, SBOM & I & I & I & A & R & I & I & I & C & I
\\\hline
Human in the loop gates and GOA kill switch thresholds & A & C & R & C & C & I & C & I & C & C
\\\hline
Privacy impact assessment and records [EU AI Act Art 9] & I & C & I & C & I & C & I & R & A & I
\\\hline
ETHOS ledger mirroring and DAO gating policy & A & I & C & R & C & I & I & C & I & I
\\\hline
Incident response playbooks across agent tiers & I & C & C & R & C & I & I & C & A & C
\\\hline
Governance KPIs and reporting cadence & A & C & C & C & I & C & I & C & R & I
\\\hline
\end{tabular}
}

\vspace{0.2cm}
\textit{Notes:} Govern controls and foundations are defined in §3.1, including SLSA, Istio, OPA/Rego, and optional on-chain accountability.
\end{table*}

\begin{table*}[!htb]
\centering
\caption{\textbf{RACIs by NIST AI RMF Function} (\textbf{Map}). Legend: R = Responsible, A = Accountable, C = Consulted, I = Informed. Roles: Exec, PO, GOA, Sec, SRE, Data, ML, Privacy, GRC, Red.}
\footnotesize
\resizebox{\textwidth}{!}{
\fontsize{11}{11}\selectfont
\begin{tabular}{|m{5.5cm}|m{1cm}|m{1cm}|m{1cm}|m{1cm}|m{1cm}|m{1cm}|m{1cm}|m{1.3cm}|m{1cm}|m{1cm}|}\hline
\textbf{Activity} & \textbf{Exec} & \textbf{PO} & \textbf{GOA} & \textbf{Sec} & \textbf{SRE} & \textbf{Data} & \textbf{ML} & \textbf{Privacy} & \textbf{GRC} & \textbf{Red} 
\\\hline
Maintain ANS registry and VC issuance & I & I & C & A & R & I & I & I & I & I
\\\hline
Keep MAESTRO threat map current per component & I & C & C & A & C & I & C & I & I & R
\\\hline
Configure Tool Gateway scopes, allow lists, rate limits & I & I & C & R & C & R & C & I & I & C
\\\hline
Inventory data sources; tag PII; RAG guardrails & I & C & I & C & I & R & C & A & C & I
\\\hline
MCP/A2A server allow listing and trust policy & I & I & C & R & C & I & C & I & I & C
\\\hline
Baseline UEBA features; log routing to PromGraf/Loki & I & I & C & A & R & C & I & I & I & C
\\\hline
Validate network policies in Istio/Cilium & I & I & C & A & R & I & I & I & I & C
\\\hline
Document system context for prioritized use cases & I & R & C & R & C & I & C & C & A & I
\\\hline
Third party tool and supplier risk entries & I & C & I & C & I & I & C & I & R & I
\\\hline
Keep Figure 2 threat-to-control crosswalk current & I & I & C & R & C & I & I & I & I & C
\\\hline
\end{tabular}
}

\vspace{0.2cm}
\textit{Notes:} ANS and Tool Gateway are the core mapping controls; Figure 2 shows the MAESTRO crosswalk.
\end{table*}

\begin{table*}[!htb]
\centering
\caption{\textbf{RACIs by NIST AI RMF Function} (\textbf{Measure}). Legend: R = Responsible, A = Accountable, C = Consulted, I = Informed. Roles: Exec, PO, GOA, Sec, SRE, Data, ML, Privacy, GRC, Red.}
\footnotesize
\resizebox{\textwidth}{!}{
\fontsize{11}{11}\selectfont
\begin{tabular}{|m{5.5cm}|m{1cm}|m{1cm}|m{1cm}|m{1cm}|m{1cm}|m{1cm}|m{1cm}|m{1.3cm}|m{1cm}|m{1cm}|}\hline
\textbf{Activity} & \textbf{Exec} & \textbf{PO} & \textbf{GOA} & \textbf{Sec} & \textbf{SRE} & \textbf{Data} & \textbf{ML} & \textbf{Privacy} & \textbf{GRC} & \textbf{Red} 
\\\hline
Calibrate UEBA anomaly thresholds and drift signals & I & I & C & A & C & C & C & I & I & R
\\\hline
Configure AIVSS scoring pipeline and factor weights & I & I & C & R & C & C & C & A & C & 
\\\hline
SSVC decision tree inputs and outcomes in GOA & I & I & C & C & C & C & C & A & C & 
\\\hline
Model evals and red team test coverage & I & I & C & C & C & I & R & I & I & A
\\\hline
PII leakage detection tests and evidence capture & I & I & I & R & C & C & A & C & C & 
\\\hline
Governance SLOs and dashboards in PromGraf & I & C & R & C & C & I & C & A & I & 
\\\hline
Rule and model drift monitoring cadence & I & I & C & A & R & C & I & C & C & 
\\\hline
Incident schema for scored events from Compliance Agent & I & I & A & R & C & C & I & C & I & 
\\\hline
Measurement review with risk posture updates & A & C & C & C & C & I & C & R & I & 
\\\hline
ZK compliance proof parameters and frequency & I & I & I & R & C & I & I & A & I & 
\\\hline
\end{tabular}
}

\vspace{0.2cm}
\textit{Notes:} AAGATE feeds AIVSS scores and uses an SSVC-style logic in the GOA’s Incident Broker. See AIVSS reference and schema in the attached doc; GOA decision flow in §3.3. 
\end{table*}

\begin{table*}[!h]
\centering
\caption{\textbf{RACIs by NIST AI RMF Function} (\textbf{Manage}). Legend: R = Responsible, A = Accountable, C = Consulted, I = Informed. Roles: Exec, PO, GOA, Sec, SRE, Data, ML, Privacy, GRC, Red.}
\footnotesize
\resizebox{\textwidth}{!}{
\fontsize{11}{11}\selectfont
\begin{tabular}{|m{5.5cm}|m{1cm}|m{1cm}|m{1cm}|m{1cm}|m{1cm}|m{1cm}|m{1cm}|m{1.3cm}|m{1cm}|m{1cm}|}\hline
\textbf{Activity} & \textbf{Exec} & \textbf{PO} & \textbf{GOA} & \textbf{Sec} & \textbf{SRE} & \textbf{Data} & \textbf{ML} & \textbf{Privacy} & \textbf{GRC} & \textbf{Red} 
\\\hline
Deploy Janus SMA per tier; define alerts & I & I & A & R & C & I & I & I & C & I
\\\hline
Triage and execute kill switch per policy & I & I & R & C & R & C & I & C & C & C
\\\hline
Quarantine via Istio AuthorizationPolicy & I & I & A & C & R & I & I & I & C & I
\\\hline
Stakeholder comms and legal notifications & A & C & C & R & C & I & C & R & R & I
\\\hline
Post-incident review; control updates & I & C & C & R & C & C & I & C & A & C
\\\hline
On-chain posting of compliance proofs & I & I & R & C & C & I & I & I & C & I
\\\hline
Risk acceptance or backlog remediation decision & A & C & C & R & C & I & C & I & R & I
\\\hline
Threat hunting and purple team exercises & I & I & A & C & A & I & I & I & C & R
\\\hline
Restore service and verify blast radius limits & I & I & C & A & R & I & I & C & C & C
\\\hline
Update allow lists and policy rules after lessons & I & I & C & R & C & C & I & C & C & I
\\\hline
\end{tabular}
}

\vspace{0.2cm}
\textit{Notes:} Janus and the millisecond kill switch are described in §§3.3–3.4; ZK proof cadence is in §4.
\end{table*}

\end{document}